\documentclass[english,letterpaper,aps,prl,twocolumn]{revtex4}
\usepackage{lmodern}
\usepackage[T1]{fontenc}
\usepackage[latin1]{inputenc}
\usepackage{bm}
\usepackage{amsmath}
\usepackage{amssymb}
\usepackage{graphicx}

\makeatletter
\@ifundefined{textcolor}{}
{%
 \definecolor{BLACK}{gray}{0}
 \definecolor{WHITE}{gray}{1}
 \definecolor{RED}{rgb}{1,0,0}
 \definecolor{GREEN}{rgb}{0,1,0}
 \definecolor{BLUE}{rgb}{0,0,1}
 \definecolor{CYAN}{cmyk}{1,0,0,0}
 \definecolor{MAGENTA}{cmyk}{0,1,0,0}
 \definecolor{YELLOW}{cmyk}{0,0,1,0}
}

\usepackage{lmodern}

\makeatletter

\usepackage{lmodern}

\makeatletter
\usepackage{lmodern}

\makeatletter

\makeatletter

\usepackage{epsfig}

\usepackage{dcolumn}

\usepackage{bm}

\usepackage{bbm}

\usepackage{wasysym}

\usepackage{marvosym}

\usepackage{subfloat}% need for subequations
\usepackage{color}% use if color is used in text
\usepackage{subfigure}% use for side-by-side figures

\usepackage{blindtext}

\newlength{\textwidthm}

\makeatother

\makeatother

\usepackage{tikz}

\makeatother

\makeatother

\makeatother

\usepackage{babel}
\begin{document}
\title{Optical selection rules of topological excitons in flat bands }
\author{Mara Lozano, Hong-Yi Xie and Bruno Uchoa$^{*}$}
\affiliation{$^{1}$Department of Physics and Astronomy, Center for Quantum Research
and Technology, University of Oklahoma, Norman, OK 73069, USA}
\email{uchoa@ou.edu}

\date{\today}
\begin{abstract}
Topological excitons are superpositions of electron-hole pair states, characterized by an envelope function with finite vorticity in momentum space. This vorticity is determined by the underlying topology of the electronic bands. We derive the optical selection rules for topological excitons in flat bands, considering different topological two-band models: a family of Hamiltonians with skyrmion pseudo-spin textures, the flattened BHZ model for a single spin and the flattened Haldane model. We derive the selection rules for these three models accounting for short-range interactions. We also consider the non-hydrogenic spectrum of excitons in the single-spin flattened BHZ model with Coulomb interactions. We show that for the case of two flat bands with skyrmion pseudo-spin textures, all excitons are bright, and the handedness of the light that couples to them is fixed by the vorticity of the pseudo-spin texture. For the single-spin flattened BHZ model, we show that bright excitons couple to circularly polarized light, regardless of the range of the interactions. In the flattened Haldane model,  topological excitons couple to elliptically polarized light. We obtain the phase diagram for the polarization of light in this model as a function of microscopic parameters of the Hamiltonian. Our results demonstrate how band topology affects exciton properties, offering a framework for predicting light-matter interactions in topological materials with flat bands.
\end{abstract}
\maketitle

\section{Introduction}

\begin{figure}[b]
\begin{centering}
\includegraphics[scale=0.23]{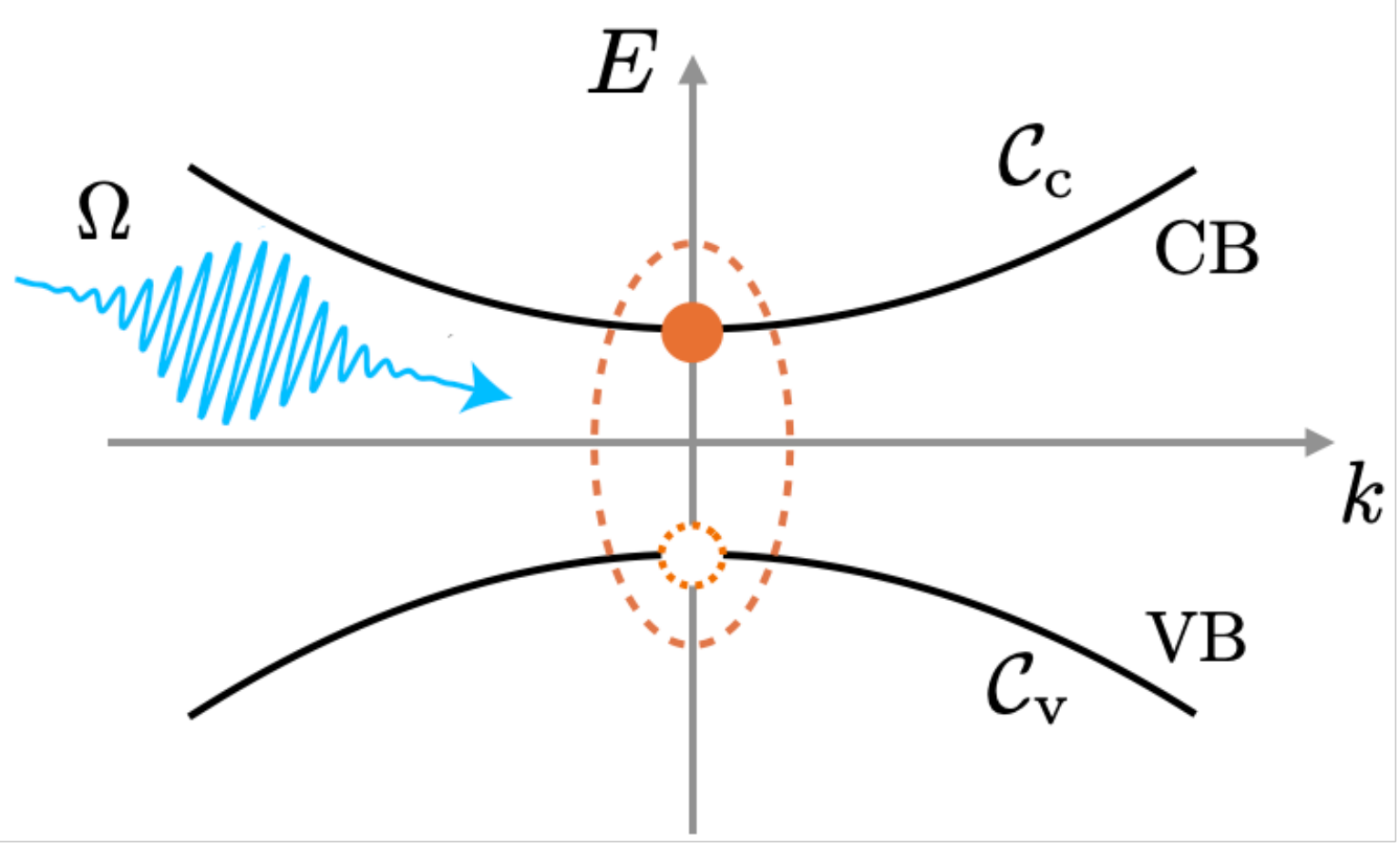}
\par\end{centering}
\caption{{\small Schematic representation of a generic topological band around
the center of the BZ in the presence of linearly polarized pumped light with frequency
$\Omega$. Topological excitons are superpositions of electron-hole
pair states, spanning the entire 2D BZ, that can emerge in flat bands
under monochromatic light. Their envelope function
has a finite vorticity in the relative momentum coordinates that is determined by the difference between the Chern numbers of the conduction
and valence bands.}}  \label{fig1}
\end{figure}

Excitons are neutral quasiparticles that form in insulators when electrons excited to the conduction band bind with the holes left behind in the valence band. In the regime where the binding energy is small compared to the insulating gap, excitons can be stabilized out of equilibrium through light pumping, as illustrated in Fig.~\ref{fig1}. Because of the $1/r$ decay of the Coulomb interaction, the spectrum of excitons in conventional insulators matches the hydrogenic Rydberg series in both two and three dimensions~\cite{Kazimierczuk2014}. More recently, signatures of quantum geometry have been identified in the non-hydrogenic spectrum of excitons in transition-metal dichalcogenide monolayers~\cite{Srivastava2015,Zhou2015,Wang2019a}, where the valleys are chiral and can be selectively excited with polarized light~\cite{Mak2012,Zeng,Cao,Mak2018}. Quantum-geometric effects were found to split the $2p$ energy levels, mimicking the effect of a Lamb shift~\cite{Srivastava2015}.

Optical selection rules for excitons are based on parity selection
rules for the interband polarization. Conventional excitons with $s$-wave
symmetry are visible when the interband polarization at the band edge
is finite~\cite{Elliot}, whereas higher angular momentum excitons
are dark and do not couple to light. Anomalous optical selection rules
have been found for insulators with gapped chiral Dirac fermions~\cite{Xiao1, Wang-1, Zhang,Cao-1,Gong},
such as gapped bilayer graphene and dichalcogenide monolayers. In
the former, $p$-wave excitons are bright and even angular momentum
states are dark~\cite{Park, Ju}. In the latter, $s$- and $d$-wave excitons
are bright, coupling to light with opposite circular polarizations~\cite{Gong}. Optical selection rules are modified in this class of
materials by the winding number in the valleys~\cite{Zhang,Cao-1},
which contributes to the angular momentum of the interband polarization
around the band edge. In all cases, the known optical selection rules
rely on the effective mass approximation, where excitons are treated
as an electron-hole pair quasiparticle with well-defined momentum
near the edge of the band. We posit that new optical selection rules
are needed in flat bands, in particular for topological excitons~\cite{Xie},
where selection rules are governed by the quantum geometric tensor
and rely instead on global properties of the Bloch bands in the Brillouin
zone (BZ). 

Topological excitons are a type of exciton that can exist when the
conduction and valence bands are topologically distinct \cite{Xie}.
The generic ket state of an exciton with center-of-mass momentum $\mathbf{Q}$
is 
\begin{equation}
|X,\mathbf{Q}\rangle=\sum_{\mathbf{k}\in\text{BZ}}\mathcal{R}_{\mathbf{k}}(\mathbf{Q})|\text{c},\mathbf{k}+\mathbf{Q}/2\rangle|\text{v},\mathbf{k}-\mathbf{Q}/2\rangle^{*}\label{eq:X}
\end{equation}
where $\mathcal{R}_{\mathbf{k}}(\mathbf{Q})$ is the envelope function
with relative momentum coordinates $\mathbf{k}$. The vorticity of
the envelope function of topological excitons is finite and set
by the difference between the Chern numbers of the conduction and
valence bands, $\zeta=\mathcal{C}_{\text{c}}-\mathcal{C}_{\text{v}}\in\mathbb{Z}$
\cite{Xie}. These excitons break time-reversal symmetry and are generically expected
to spontaneously emit circularly polarized light through radiative
decay. Since vorticity is a global property in a 2D BZ, topological
excitons cannot be approximated as electron-hole pairs concentrated
around individual points of high Berry curvature near the band edge.
On the contrary, their envelope function extends over the entire
BZ. Topological excitons have been generically predicted to exist in Chern bands driven with pumped light  \cite{Xie}. Excitons with non-trivial topology were also predicted in light pumped organic semiconductors in one dimension \cite{Jankowski, Thompson}, and in the ground state of moire heterostructures \cite{Wu2,Xie2, Kwan}, monolayer transition metal dichalchogenides \cite{Gong2}, quantum spin Hall systems \cite{Chen2,Blasson},  and one dimensional topological crystals \cite{Davenport}.

Short range interactions generate up to three midgap bound states
of topological excitons in flat bands. This contrasts with conventional
excitons, where only a single bound state is expected. Even though Coulomb
interactions produce a full energy spectrum with an infinite number
of bound states, quantum-geometric effects \cite{Srivastava2015,Zhou2015,Wu}
and topological constraints are expected to render the excitonic energy
spectrum highly non-hydrogenic. 

In this paper, we address in detail the optical selection rules for
the coupling of topological excitons with light. We express the exciton
oscillation strength in the presence of light in terms of the envelope function and then calculate the effective exciton dipole moment $\boldsymbol{\ell} $, which
determines the selection rules in flat bands. We first examine the case of contact
interactions, where general analytical expressions for the effective
exciton dipole moment can be calculated {\it exactly}, and then consider
various examples of two-band topological models: a family of Hamiltonians obeying
a minimal-energy configuration of the non-linear sigma model (NLsM),
with a skyrmion pseudo-spin texture, the flattened BHZ model~\cite{Bernevig} for a single spin, which will be referred to simply as `flattened BHZ model' for brevity, 
and the flattened Haldane model~\cite{Haldane}. We then examine the
case of Coulomb interactions for the flattened BHZ model. 

For the case of contact interactions, we show that any flat two-band
topological Hamiltonian with a skyrmion pseudospin texture produces
topological excitons that couple to light with a specific circular polarization, which is determined by the sign of the Chern number of the
valence band. Both the flattened BHZ model and the flattened Haldane model
generate three topological exciton bound states whose envelope functions
have vorticity $\zeta=2$. We show that two of these excitons are bright
and one dark. In particular, the bright excitons in the flattened
BHZ model couple always to circularly polarized light with opposite
polarizations, with the handedness of the light polarization for each
exciton being predetermined by the sign of the Chern number of the
valence band. In the flattened Haldane model the bright topological
excitons couple to elliptically polarized light, also with opposite
polarizations. Using a suitable parametrization for the polarization
of the light with Jones vectors, we determine the phase diagram for
the light polarization according to the optical selection rules, as
a function of the microscopic parameters of the Hamiltonian. 

We finally address the Coulomb case in the flattened BHZ model, where
an infinite number of excitonic bound states emerge, each one carrying an envelope function with the same vorticity
$\zeta=2$. The oscillation strength of the higher energy exciton
states decay exponentially with increasing $\nu$, and therefore these
states are much less bright than the one in the lowest energy state. No dark excitons
are present in the non-hydrogenic spectrum of topological excitons
for the chosen model, with all excitons coupling to circularly polarized
light. The selection rules determine the handedness of the light polarization
that couples to each of the exciton states. 

The paper is organized as follows: in Sec.~\ref{sec-2} we derive the dielectric
response of generic flat two-band Hamiltonians and their exciton oscillation
strength in the presence of a pumped light field, and determine their
general optical selection rules. In Sec.~\ref{sec-3} we consider the case
of topological excitons in three different flat-band
models. Finally, in Sec.~\ref{sec-4} we present our discussion. 

\section{Dielectric response of flat bands} \label{sec-2}

We assume a generic 2D physical system formed by two flat bands in
the presence of monochromatic light pumping. The flat bands have Chern
numbers $\mathcal{C}_{\text{c}}$ and $\mathcal{C}_{\text{v}}$ for
the conduction and valence bands, respectively, separated by an energy
gap $\Delta$. We define the periodic part of the Bloch wavefunctions
as $|n,\mathbf{k}\rangle$, where $n=\text{c},\text{v}$ labels the
bands with momentum $\mathbf{k}$. 

Monochromatic photons can coherently induce population inversion between
flat bands over the whole BZ. The incident monochromatic,  linearly polarized, light field
is given by $\textbf{E}(t)=\textbf{E}e^{-i\Omega t}+\textbf{E}^{*}e^{i\Omega t},$
where $\textbf{E}=E(\cos\theta\,\hat{\textbf{x}}+i\sin\theta\,\hat{\textbf{y}})$
defines the complex amplitude of the electric field, with $E$, $\Omega$,
and $\theta$ denoting the amplitude, frequency, and polarization
angle of the light, respectively. Within the rotating-wave approximation,
the energy-basis time-dependent Hamiltonian of a two-band insulator
driven by monochromatic light is \cite{Xie}
\begin{equation}
\hat{\mathcal{H}}_{\textbf{k}}(t)=\begin{pmatrix}\varepsilon_{\mathrm{c}} & e^{-i\Omega t}\textbf{E}\cdot\textbf{P}_{\textbf{k}}\\
e^{i\Omega t}\textbf{E}^{*}\cdot\textbf{P}_{\textbf{k}}^{*} & \varepsilon_{\mathrm{v}}
\end{pmatrix},\label{eq:2band}
\end{equation}
where $\varepsilon_{\text{c,v}}$ give the energies of the  conduction
and valence bands in the flat band limit, $\varepsilon_{\text{c}}-\varepsilon_{\text{v}}=\Delta$.
$\mathbf{P}_{\mathbf{k}}$ is the electric polarization in linear
response, 
\begin{equation}
\mathbf{P}_{\mathbf{k}}=\frac{e\Delta}{\Omega}\boldsymbol{\mathcal{A}}_{\mathrm{cv}}(\mathbf{k})+\mathbf{P}_{\text{ex},\mathbf{k}}.\label{eq:P-2}
\end{equation}
where $\boldsymbol{\mathcal{A}}_{\mathrm{cv}}(\mathbf{k})=-i\langle\mathrm{c},\textbf{k}\rvert\partial_{\textbf{k}}\lvert\mathrm{v},\textbf{k}\rangle$
is the interband Berry connection, and 
\begin{equation}
\textbf{P}_{\text{ex},\textbf{k}}=\sum_{\nu}\frac{\varepsilon_{\mathrm{B},\nu}\mathcal{R}_{\nu,\textbf{k}}\boldsymbol{\ell}_{\nu}}{\Omega-\varepsilon_{\nu}+i\gamma}\label{eq:Pex}
\end{equation}
is the polarization of electron-hole pairs, which gives the exciton
contribution to the total electric polarization for flat bands. $\mathcal{R}_{\nu,\mathbf{k}}$
is the envelope function of exciton mode $\nu$ with center of mass
momentum $\mathbf{Q}=0$ and energy $\varepsilon_{\nu}$, $\varepsilon_{\mathrm{B},\nu}=\Delta-\varepsilon_{\nu}$
is the exciton binding energy, $\gamma$ is the decay rate due to
the bath, and
\begin{equation}
\boldsymbol{\ell}_{\nu}=\frac{e\Delta}{\Omega}\sum_{\textbf{k}\in\text{BZ}}\mathcal{R}_{\nu,\textbf{k}}^{*}\boldsymbol{\mathcal{A}}_{\mathrm{cv}}(\mathbf{k})\label{eq:ell-1}
\end{equation}
plays a role of the effective exciton dipole moment that couples
to light. As it will become clear, this quantity determines the optical
selection rules for excitons in flat bands.

The second term in the electric polarization (\ref{eq:P-2}) results
from the interaction in the excitonic channels. The first term follows
from the quantum-geometric contribution to the electric polarization.
This can be seen from a generic Hamiltonian in the orbital basis,
which can be written as
\begin{equation}
\hat{\mathcal{K}}_{\mathbf{k}}=\hat{U}(\mathbf{k})\hat{\varepsilon}(\mathbf{k})\hat{U}^{\dagger}(\mathbf{k}),\label{eq:K}
\end{equation}
where $\hat{\varepsilon}(\mathbf{k})\equiv\text{diag}\{\varepsilon_{n}(\mathbf{k})\}$
and $\hat{U}(\mathbf{k})$ is the unitary transformation that relates
the original orbital basis $|i,\mathbf{k}\rangle$, with $i=1,2$,
to the energy basis, $|n,\mathbf{k}\rangle=\sum_{i=1,2}U_{n,i}|i,\mathbf{k}\rangle.$
The time-dependent Hamiltonian in the presence of the light field
can be expressed in the energy basis as $\hat{\mathcal{K}}_{\mathbf{k}}^{\text{d}}(t)=\hat{U}^{\dagger}(\mathbf{k})\hat{\mathcal{K}}_{\mathbf{k}-e\mathbf{A}(t)}\hat{U}(\mathbf{k})$,
namely
\begin{equation}
\hat{\mathcal{K}}_{\mathbf{k}}^{\text{d}}(t)=\hat{\varepsilon}(\mathbf{k})+\mathbf{A}(t)\cdot\hat{\boldsymbol{v}}_{\mathbf{k}}^{\text{d}}+\mathcal{O}(\mathbf{A}^{2}),\label{eq:=000020expansion}
\end{equation}
where $\hat{\boldsymbol{v}}_{\mathbf{k}}^{\text{d}}=\partial_{\mathbf{k}}\hat{\varepsilon}(\mathbf{k})+[\hat{\bm{\mathcal{A}}}(\mathbf{k}),\hat{\varepsilon}(\mathbf{k})]$
is the velocity operator of the flat band quasiparticles in the energy
basis, which has an anomalous contribution due to quantum geometry
\cite{Wen}, $\mathbf{A}(t)=-i(\textbf{E}e^{-i\Omega t}-\textbf{E}^{*}e^{i\Omega t})/\Omega$
is the vector potential and
\begin{equation}
\hat{\bm{\mathcal{A}}}(\mathbf{k})=\hat{U}^{\dagger}(\mathbf{k})\partial_{\mathbf{k}}\hat{U}(\mathbf{k})\label{eq:At}
\end{equation}
is the Berry connection tensor. For the purposes of this paper, we
assume the flat band limit, where the group velocity $\partial_{\mathbf{k}}\hat{\varepsilon}(\mathbf{k})$
is small compared to the anomalous contribution to the velocity of
the quasiparticles. In that limit, the energy of the flat bands is
unaffected by the external electric field. 

The total electric dipole moment per unit cell is 
\begin{equation}
\boldsymbol{\mathcal{P}}(t)\equiv\sum_{\textbf{k}\in\text{BZ}}\textbf{P}_{\textbf{k}}^{*}\rho_{\text{cv},\mathbf{k}}(t,t)+\mathrm{c.c.},\label{eq:dipole}
\end{equation}
where 
\begin{equation}
\rho_{\text{cv},\mathbf{k}}(t,t)=-e^{-i\Omega t}\delta f\frac{e\Delta}{\Omega}\frac{\textbf{E}\cdot\boldsymbol{\mathcal{A}}_{\mathrm{cv}}(\mathbf{k})}{\Omega-\Delta+i\gamma}\label{eq:rho}
\end{equation}
is the interband density matrix up to linear order in the electric
field \cite{Haug}, with $\delta f=f(\varepsilon_{\text{v}})-f(\varepsilon_{\text{c}})$
set by the difference of Fermi distributions between conduction and
valence bands. In the frequency representation, one can explicitly
relate the electric dipole
\begin{equation}
\boldsymbol{\mathcal{P}}(\Omega)=\sum_{\beta}\chi_{\alpha\beta}(\Omega)E_{\beta}(\Omega)\label{eq:P3}
\end{equation}
to the susceptibility per unit cell $\chi_{\alpha\beta}$, with $\alpha,\beta=x,y$.
Substituting Eq. (\ref{eq:P-2}) and (\ref{eq:ell-1}) into (\ref{eq:dipole}),
the susceptibility can be decomposed as 
\begin{equation}
\chi_{\alpha\beta}(\Omega)=\chi_{\mathrm{cv},\alpha\beta}(\Omega)+\sum_{\nu}\chi_{\nu,\alpha\beta}(\Omega),\label{eq:chi2}
\end{equation}
where
\begin{equation}
\chi_{\text{cv},\alpha\beta}(\Omega)=-\delta f\left(\frac{\theta(\Omega)O_{\text{cv},\alpha\beta}}{\Omega-\Delta+i\gamma}-\frac{\theta(-\Omega)O_{\text{cv},\alpha\beta}^{*}}{\Omega+\Delta+i\gamma}\right)\label{eq:Chicv}
\end{equation}
and 
\begin{equation}
\chi_{\nu,\alpha\beta}(\Omega)=-\delta f\left(\frac{\theta(\Omega)O_{\nu,\alpha\beta}}{\Omega-\varepsilon_{\nu}+i\gamma}-\frac{\theta(-\Omega)O_{\nu,\alpha\beta}^{*}}{\Omega+\varepsilon_{\nu}+i\gamma}\right),\label{eq:Chinu}
\end{equation}
with $\theta(\Omega)$ the Heaviside step function. $\chi_{\mathrm{cv},\alpha\beta}(\Omega)$
and $\chi_{\nu,\alpha\beta}(\Omega)$ are the interband and excitonic
contributions respectively, with 
\begin{equation}
O_{\mathrm{cv},\alpha\beta}\equiv\left(\frac{e\Delta}{\Omega}\right)^{2}\sum_{\textbf{k}\in\text{BZ}}\mathcal{A}_{\mathrm{vc},\alpha}(\mathbf{k})\mathcal{A}_{\mathrm{cv},\beta}(\mathbf{k}),\label{eq:Theta0}
\end{equation}
and 
\begin{equation}
O_{\nu,\alpha\beta}\equiv\ell_{\nu,\alpha}^{*}\ell_{\nu,\beta}.\label{eq:theta}
\end{equation}
$\hat{O}_{\mathrm{cv}}$ and $\hat{O}_{\nu}$ are $2\times2$ Hermitian
matrices whose eigenvalues determine the oscillation strength of particle-hole
excitations and excitons, respectively, in the presence of pumped
light. 

\subsection{Optical selection rules} 

At the exciton resonant condition $\Omega\approx\pm\varepsilon_{\nu}$
the susceptibility is dominated by the exciton contribution. For non-degenerate
exciton bands, the eigenproblem is 
\begin{equation}
\hat{O}_{\nu}\boldsymbol{\xi}_{\nu,\sigma}=o_{\nu,\sigma}\boldsymbol{\xi}_{\nu,\sigma},
\end{equation}
with the orthonormality condition $\boldsymbol{\xi}_{\nu,\sigma}^{*}\cdot\boldsymbol{\xi}_{\nu,\sigma'}=\delta_{\sigma\sigma'}$.
The eigenvalues are 
\begin{equation}
o_{\nu,\sigma=\pm1}=\frac{1+\sigma}{2}|\boldsymbol{\ell}_{\nu}|^{2},\label{ExEiVal}
\end{equation}
with the eigenvectors 
\begin{equation}
\boldsymbol{\xi}_{\nu,+}=\begin{pmatrix}\hat{\ell}_{\nu,x}^{*}\\
\hat{\ell}_{\nu,y}^{*}
\end{pmatrix},\quad\boldsymbol{\xi}_{\nu,-}=\begin{pmatrix}-\hat{\ell}_{\nu,y}\\
\hat{\ell}_{\nu,x}
\end{pmatrix},\label{SelecRule}
\end{equation}
where $\hat{\boldsymbol{\ell}}_{\nu}\equiv\boldsymbol{\ell}_{\nu}/|\boldsymbol{\ell}_{\nu}|$
is a unit vector. 

The optical dielectric function of the system is
\begin{equation}
\epsilon_{\alpha\beta}(\Omega)=\epsilon_{0}+\chi_{\alpha\beta}(\Omega),\label{eq:epsilon}
\end{equation}
with $\epsilon_{0}$ the vacuum permittivity. Substituting into the
homogeneous frequency-dependent Maxwell equation for light propagating
along the $z$-axis, 
\begin{equation}
\partial_{z}^{2}E_{\alpha}(z,\Omega)+\frac{\Omega^{2}}{c^{2}}\epsilon_{\alpha\beta}(\Omega)E_{\beta}(z,\Omega)=0,\label{MaxEq}
\end{equation}
where $c$ is the speed of light, we find two distinct modes. The exciton contribution to the dielectric
function vanishes in the $\boldsymbol{\xi}_{\nu,-}$ mode ($o_{\nu,\sigma=-1}=0$).
This eigenmode is thus transparent to light and corresponds to a dark
mode. In contrast, the $\boldsymbol{\xi}_{\nu,+}$ eigenmode is bright
whenever $|\boldsymbol{\ell}_{\nu}|\neq0$, with dielectric function
\begin{equation}
\epsilon_{\nu}(\Omega)=\epsilon_{0}-\frac{|\boldsymbol{\ell}_{\nu}|^{2}\delta f}{\Omega-\varepsilon_{\nu}+i\gamma}.\label{eq:epsilon2}
\end{equation}
This mode has the absorption coefficient 
\begin{equation}
\alpha_{\nu}(\Omega)=\frac{\Omega}{n_{\nu}(\Omega)c}\text{Im}\epsilon_{\nu}(\Omega),\label{eq:alpha}
\end{equation}
where $n_{\nu}(\Omega)=\sqrt{\frac{1}{2}\left(\text{Re}\left[\epsilon_{\nu}(\Omega)\right]+\left|\epsilon_{\nu}(\Omega)\right|\right)}$
is the index of refraction. 

From Eq. (\ref{SelecRule}) it becomes clear that the vector $\boldsymbol{\ell}_{\nu}=\boldsymbol{\xi}_{\nu,+}^{*}$
defines the optical selection rule in flat bands. The exciton bound
state $\nu$ couples to the light field if and only if $\textbf{E}\cdot\boldsymbol{\ell}_{\nu}\neq0$.
The quantity 
\begin{equation}
o_{\nu,\sigma=+1}=|\boldsymbol{\ell}_{\nu}|^{2}\label{eq:OS}
\end{equation}
is the exciton oscillation strength, which determines its brightness. 

Near the absorption edge $\Omega\approx\pm\Delta$, interband processes
dominate the susceptibility and the dielectric response. The eigenproblem
for interband electron-hole excitations in this frequency regime is
defined by 
\begin{equation}
\hat{O}_{\mathrm{cv}}\boldsymbol{\eta}_{\mathrm{cv},\sigma}=o_{\mathrm{cv},\sigma}\boldsymbol{\eta}_{\mathrm{cv},\sigma},
\end{equation}
with the orthonormality condition $\boldsymbol{\eta}_{\mathrm{cv},\sigma}^{*}\cdot\boldsymbol{\eta}_{\mathrm{cv},\sigma'}=\delta_{\sigma\sigma'}$.

By inspection, one can calculate the eigenvectors of $\hat{O}_{\mathrm{cv}}$
for the flattened model of massive Dirac fermions. This continuum
model is described by the two band Hamiltonian in the orbital basis,
$\hat{\mathcal{K}}_{\textbf{k}}=\frac{\Delta}{2}\hat{\textbf{d}}(\textbf{k})\cdot\boldsymbol{\sigma},$
with $\boldsymbol{\sigma}=(\sigma_{1},\sigma_{2},\sigma_{3})$ a vector
of Pauli matrices and $\hat{\textbf{d}}(\textbf{k})=\textbf{d}(\textbf{k})/|\textbf{d}(\textbf{k})|$
a unit vector defined by $d_{1}(\mathbf{k})=v_{F}k_{x}$, $d_{2}(\mathbf{k})=\tau v_{F}k_{y}$,
and $d_{3}(\mathbf{k})=\Delta$. $v_{F}$ is the Fermi velocity, and
$\tau=\pm1$ denotes the valleys, which determine the chirality of
the Dirac fermions. At the Dirac point $\textbf{k}=0$, the interband
Berry connection satisfies 
\begin{equation}
\mathcal{A}_{\mathrm{cv},x}(0)=i\tau\mathcal{A}_{\mathrm{cv},y}(0)=\frac{1}{2k_{0}},\label{eq:A-1}
\end{equation}
where $k_{0}\equiv\Delta_{0}/v_{F}>0$. Substitution of Eq. (\ref{eq:A-1})
into Eq.~(\ref{eq:Theta0}) yields 
\begin{equation}
o_{\mathrm{cv},\sigma=\pm1}=\frac{e^{2}(1+\sigma)}{4k_{0}^{2}},
\end{equation}
\begin{equation}
\boldsymbol{\eta}_{+}=\frac{1}{\sqrt{2}}\begin{pmatrix}1\\
i\tau
\end{pmatrix},\quad\boldsymbol{\eta}_{-}=\frac{1}{\sqrt{2}}\begin{pmatrix}-1\\
i\tau
\end{pmatrix}.
\end{equation}
The $\sigma=-1$ mode has zero eigenvalue, $o_{\mathrm{cv},\sigma=-1}=0$,
and is thus transparent to light. On the other hand, the $\boldsymbol{\eta}_{\sigma=+1}$
eigenmode selectively couples to circularly polarized light
with the same polarization $\boldsymbol{\eta}_{+}$. Therefore,
the $\tau=1$ valley will couple with one handedness of light polarization
and the $\tau=-1$ valley with the opposite handedness,  in agreement with standard optical 
selection rules for Dirac fermions \cite{Zhang,Cao-1}.

\section{Topological excitons} \label{sec-3}

The exciton envelope functions in flat bands satisfy
the Bethe-Salpeter equation
\begin{equation}
\sum_{\mathbf{q}^{\prime}\in\text{BZ}}h_{\mathbf{q}\mathbf{q}^{\prime}}(\mathbf{Q})\mathcal{R}_{\nu,\mathbf{q}^{\prime}}(\mathbf{Q})=\varepsilon_{\nu}(\mathbf{Q})\mathcal{R}_{\nu,\mathbf{q}}(\mathbf{Q}),\label{eq:Wannier-1}
\end{equation}
also known as the Wannier equation, where 
\begin{equation}
h_{\mathbf{q}\mathbf{q}^{\prime}}(\mathbf{Q})=\delta_{\textbf{q},\textbf{q}^{\prime}}\Delta-\delta f\,W_{\textbf{q},\textbf{q}^{\prime};\textbf{Q}}, \label{eq:ExcHam-1}
\end{equation}
is the exciton Hamiltonian. The envelope functions are orthonormal $\sum_{\textbf{k}\in\text{BZ}}\mathcal{R}_{\nu,\textbf{k}}^{*}\mathcal{R}_{\nu^{\prime},\textbf{k}}=\delta_{\nu\nu^{\prime}}$. The solution of the Wannier equation (\ref{eq:Wannier-1})
gives the energies of the excitons as functions of the center-of-mass
momentum $\mathbf{Q}$ and their envelope functions. The interaction
coefficient $W_{\mathbf{q},\mathbf{q}^{\prime};\mathbf{Q}}=W_{\mathbf{q},\mathbf{q}^{\prime};\mathbf{Q}}^{(\text{d})}-W_{\mathbf{q},\mathbf{q}^{\prime};\mathbf{Q}}^{(\text{e})}$
can be decomposed into two terms corresponding to the direct and the exchange
contributions, 
\begin{align}
W_{\mathbf{q},\mathbf{q}^{\prime};\mathbf{Q}}^{(\text{d})} & =v(\mathbf{q}-\mathbf{q}^{\prime})\,\mathcal{U}_{\mathbf{q}+\frac{\mathbf{Q}}{2},\mathbf{q}^{\prime}+\frac{\mathbf{Q}}{2}}^{\text{cc}}\mathcal{U}_{\mathbf{q}^{\prime}-\frac{\mathbf{Q}}{2},\mathbf{q}-\frac{\mathbf{Q}}{2}}^{\text{vv}}, \label{eq:Wd}\\
W_{\mathbf{q},\mathbf{q}^{\prime};\mathbf{Q}}^{(\text{e})} & =v(\mathbf{Q})\,\mathcal{U}_{\mathbf{q}+\frac{\mathbf{Q}}{2},\mathbf{q}-\frac{\mathbf{Q}}{2}}^{\text{cv}}\mathcal{U}_{\mathbf{q}^{\prime}-\frac{\mathbf{Q}}{2},\mathbf{q}^{\prime}+\frac{\mathbf{Q}}{2}}^{\text{vc}}.\label{eq:We}
\end{align}
The matrix
\begin{equation}
\hat{\mathcal{U}}_{\mathbf{k},\mathbf{q}}\equiv\hat{U}^{\dagger}(\mathbf{k})\hat{U}(\mathbf{q}), \label{eq:U}
\end{equation}
 accounts for the contribution of the quantum geometry of the bands.
Since only excitons with zero center of mass momentum couple to light,
we will set $\mathbf{Q}=0$ for now on. 

\subsection{Contact interaction}

The exact solution of the Wannier equation (\ref{eq:Wannier-1}) for
flat bands in the presence of contact interactions $v(\mathbf{q})=v$
was originally derived in Ref. \cite{Xie} under
symmetry constraints. Here we derive the general solution of the envelope function $\mathcal{R}_{\nu,\mathbf{q}}$ and employ the solution
to calculate general expressions for the polarization of topological
excitons. We then apply these results to three different models with
topological flat bands and explicitly derive their optical selection
rules. 

Let us define the envelope functions at $\mathbf{Q}=0$ satisfying
the Wannier Eq. (\ref{eq:Wannier-1}) as $\mathcal{R}_{\nu,\textbf{k}}\equiv\langle\textbf{k},0\rvert\psi_{\nu}\rangle$,
where
\begin{equation}
|\mathbf{k},0\rangle\equiv|\text{c},\mathbf{k}\rangle|\text{v},\mathbf{k}\rangle^{*}=\left(\begin{array}{c}
U_{1,\text{c},\mathbf{q}}\\
U_{2,\text{c},\mathbf{q}}
\end{array}\right)\otimes\left(\begin{array}{c}
U_{1,\text{v},\mathbf{q}}^{*}\\
U_{2,\text{v},\mathbf{q}}^{*}
\end{array}\right)\label{eq:ehproduct}
\end{equation}
denotes the e-h pair state described by a 4-component spinor in the
electron orbital basis and
\begin{equation}
W_{\mathbf{q},\mathbf{q}^{\prime};0}=v\langle\mathbf{q},0|\mathbf{q}^{\prime},0\rangle.\label{eq:W-1-1}
\end{equation}
Then $\rvert\psi_{\nu}\rangle$ must be eigenstates
of the $4\times4$ auxiliary matrix $\hat{w}$,
\begin{equation}
\hat{w}\equiv \sum_{\mathbf{q}\in\text{BZ}}|\mathbf{q},0\rangle\langle\mathbf{q},0|,\label{eq:aux-1}
\end{equation}
that satisfy the eigenvalue problem $\hat{w}\rvert\psi_{\nu}\rangle=w_{\nu}\rvert\psi_{\nu}\rangle$
and obey the orthonormality condition $\langle\psi_{\nu}\rvert\psi_{\nu^{\prime}}\rangle=\delta_{\nu\nu^{\prime}}/w_{\nu}$.
This can be immediately seen by noticing that
\begin{align}
\sum_{\mathbf{q}^{\prime}}W_{\mathbf{q},\mathbf{q}^{\prime}}\mathcal{R}_{\nu,\mathbf{q}^{\prime}} & =v\sum_{\mathbf{q}^{\prime}}\langle\mathbf{q},0|\mathbf{q},0\rangle\langle\mathbf{q},0\rvert\psi_{\nu}\rangle\nonumber \\
 & =v \langle\mathbf{q},0|\hat{w}|\psi_{\nu}\rangle=v w_{\nu}\mathcal{R}_{\mathbf{q}}.\label{eq:Id}
\end{align}
Therefore, short range interactions between electron and holes occupying
two bands permit at most four excitonic bound states. After substitution
of Eq. (\ref{eq:Id}) into the Wannier equation (\ref{eq:Wannier-1}),
the eigenvalues $w_{\nu}$ give the energy of the exciton bound states,
\begin{equation}
\varepsilon_{\nu}=\Delta-vw_{\nu}.\label{eq:binding}
\end{equation}

We now assume that the kinetic Hamiltonian of two flat bands has the
general form 
\begin{equation}
\mathcal{\hat{K}}_{\mathbf{k}}=\frac{\Delta}{2}\hat{\mathbf{d}}(\mathbf{k})\cdot\boldsymbol{\sigma},\label{eq:K2}
\end{equation}
where $\hat{\mathbf{d}}(\mathbf{k})=(\hat{d}_{1},\hat{d}_{2},\hat{d}_{3})$
is an arbitrary unit vector on the sphere. From Eq. (\ref{eq:ehproduct}),
the operator $\hat{w}$ is given by
\begin{equation}
\hat{w}\equiv\frac{1}{4}\left(1+\sum_{i=1}^{3}c_{i}\hat{\gamma}_{i}+\sum_{ij=1}^{3}c_{ij}\hat{\gamma}_{ij}\right),\label{AuxMat-1}
\end{equation}
where $\hat{\gamma}_{i}$ and $\hat{\gamma}_{ij}$ are the following
$\gamma$ matrices
\begin{align*}
\hat{\gamma}_{1}=\begin{pmatrix}0 & -1 & 1 & 0\\
-1 & 0 & 0 & 1\\
1 & 0 & 0 & -1\\
0 & 1 & -1 & 0
\end{pmatrix},\quad & \hat{\gamma}_{2}=\begin{pmatrix}0 & -i & -i & 0\\
i & 0 & 0 & -i\\
i & 0 & 0 & -i\\
0 & i & i & 0
\end{pmatrix}\\
\hat{\gamma}_{3}=\begin{pmatrix}0 & 0 & 0 & 0\\
0 & 2 & 0 & 0\\
0 & 0 & -2 & 0\\
0 & 0 & 0 & 0
\end{pmatrix},\quad & \hat{\gamma}_{11}=\begin{pmatrix}0 & 0 & 0 & -1\\
0 & 0 & -1 & 0\\
0 & -1 & 0 & 0\\
-1 & 0 & 0 & 0
\end{pmatrix}\\
\hat{\gamma}_{22}=\begin{pmatrix}0 & 0 & 0 & -1\\
0 & 0 & 1 & 0\\
0 & 1 & 0 & 0\\
-1 & 0 & 0 & 0
\end{pmatrix},\quad & \hat{\gamma}_{33}=\begin{pmatrix}-1 & 0 & 0 & 0\\
0 & 1 & 0 & 0\\
0 & 0 & 1 & 0\\
0 & 0 & 0 & -1
\end{pmatrix}\\
\hat{\gamma}_{12}=\begin{pmatrix}0 & 0 & 0 & -i\\
0 & 0 & i & 0\\
0 & -i & 0 & 0\\
i & 0 & 0 & 0
\end{pmatrix},\quad & \hat{\gamma}_{21}=\begin{pmatrix}0 & 0 & 0 & i\\
0 & 0 & i & 0\\
0 & -i & 0 & 0\\
-i & 0 & 0 & 0
\end{pmatrix}\\
\hat{\gamma}_{13}=\begin{pmatrix}0 & 0 & -1 & 0\\
0 & 0 & 0 & 1\\
-1 & 0 & 0 & 0\\
0 & 1 & 0 & 0
\end{pmatrix},\quad & \hat{\gamma}_{31}=\begin{pmatrix}0 & -1 & 0 & 0\\
-1 & 0 & 0 & 0\\
0 & 0 & 0 & 1\\
0 & 0 & 1 & 0
\end{pmatrix}\\
\hat{\gamma}_{23}=\begin{pmatrix}0 & 0 & i & 0\\
0 & 0 & 0 & -i\\
-i & 0 & 0 & 0\\
0 & i & 0 & 0
\end{pmatrix},\quad & \hat{\gamma}_{32}=\begin{pmatrix}0 & -i & 0 & 0\\
i & 0 & 0 & 0\\
0 & 0 & 0 & i\\
0 & 0 & -i & 0
\end{pmatrix},
\end{align*}
with coefficients
\begin{equation}
c_{i}=\sum_{\textbf{q}\in\text{BZ}}\hat{d}_{i}(\textbf{q})\label{eq:ci}
\end{equation}
and
\begin{equation}
c_{ij}=\sum_{\textbf{q}\in\text{BZ}}\hat{d}_{i}(\textbf{q})\hat{d}_{j}(\textbf{q}).\label{eq:cij}
\end{equation}
The auxiliary operator $\hat{w}$ in Eq.~\eqref{AuxMat-1} obeys the following properties:
\emph{i)} $c_{ij}=c_{ji}$, \emph{ii)} $\sum_{i}c_{ii}=1$, \emph{iii)}
the $\gamma$ matrices are orthogonal, i.e., $\mathrm{tr}(\hat{\gamma}_{i}\hat{\gamma}_{j})\propto\delta_{ij}$.

The matrix $\hat{w}$ has a zero eigenvalue and three non-zero ones.
Therefore, we can take the following decomposition
\begin{equation}
\hat{w}=\hat{u}\begin{pmatrix}0 & 0\\
0 & \hat{\sf{w}}
\end{pmatrix}\hat{u}^{\dagger},
\end{equation}
where
\begin{equation}
\hat{\sf{w}}_{ij}=\frac{1}{2}(\delta_{ij}-c_{ij}-i\epsilon_{ijl}c_{l}),
\end{equation}
and
\begin{equation}
\hat{u}=\frac{1}{\sqrt{2}}\begin{pmatrix}1 & 0 & 0 & 1\\
0 & 1 & -i & 0\\
0 & 1 & i & 0\\
1 & 0 & 0 & -1
\end{pmatrix},
\end{equation}
with $i,j,l\in\{1,2,3\}$, $\epsilon_{ijl}$ being the Levi-Civita
symbol and $\hat{u}$ unitary. Now diagonalizing $\hat{\sf{w}}$ through
the $3\times3$ unitary matrix $\hat{\sf{u}}$,
\begin{equation}
\hat{w}=\hat{u}\begin{pmatrix}0 & 0\\
0 & \hat{\sf{u}}
\end{pmatrix} \mathrm{diag}\{0,\, w_1, \, w_2,\, w_3\} \begin{pmatrix}0 & 0\\
0 & \hat{\sf{u}}^{\dagger}
\end{pmatrix}\hat{u}^{\dagger}\,,
\end{equation}
the eigenkets of $\hat{w}$ corresponding to excitonic bound states
are 
\begin{equation}
\psi_{\nu,j}=\frac{1}{\sqrt{w_{\nu}}} \sum_{l}u_{j,l} \, {\sf{u}}_{l ,\nu},\label{eq:psi}
\end{equation}
where $\psi_{\nu,j}$ is the $j$-th component of the eigenket $|\psi_{\nu}\rangle$. 

\begin{figure*}
\begin{centering}
\includegraphics[scale=0.47]{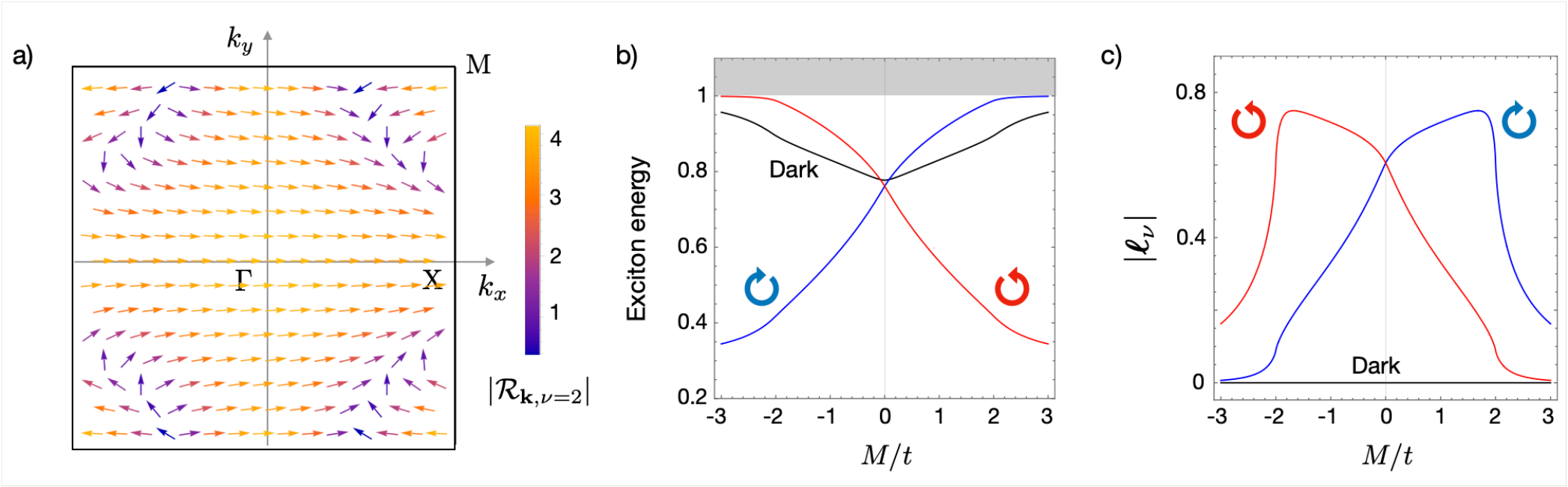}
\par\end{centering}
\caption{{\small Topological excitons in the flattened BHZ model for short range
interaction with strength $v=0.7\Delta$ and $M/t=1$. a) Profile function
$\mathcal{R}_{\nu=2,\mathbf{k}}(0)$ of the $\nu=2$ exciton bound
state in the BZ centered at the $\Gamma$ point. Arrows indicate the phase and color bar the amplitude. $\mathcal{R}_{\nu,\mathbf{k}}(0)$
($\nu=1,2,3$) has vorticity $\zeta=2$. In panels b) and c) red and
blue lines represent bright excitons $\nu =2,3$ respectively. The excitons couple to circularly polarized
light, with the polarization indicated in the circles. Black lines
represent the dark exciton $\nu=1$. b) Energy of the exciton bound states
(in units of the energy gap $\Delta$) as functions of the ratio $M/t$.
The dark gray region indicates the continuum of single particle states.
c) Effective exciton dipole moment $|\boldsymbol{\ell}_{\nu}|$ $(\nu=1,2,3)$
as functions of $M/t$ in units of $ea$.  }} \label{fig2}
\end{figure*}

Using the definition of the dipole moment given in Eq.~(\ref{eq:ell-1})
together with the eigenstates $\rvert\psi_{\nu}\rangle$, we obtain
the expression for the electric dipole moment of the three exciton
bound states 
\begin{equation}
\boldsymbol{\ell}_{\nu}=\frac{1}{4}\langle\psi_{\nu}\rvert\textbf{L}\rangle, \label{DipMomExc}
\end{equation}
with $\nu=1,2,3$, where 
\begin{equation}
\rvert\textbf{L}\rangle\equiv4e\sum_{\mathbf{k}\in\text{BZ}}\boldsymbol{\mathcal{A}}_{\text{cv}}(\mathbf{k})|\mathbf{k},0\rangle\label{eq:Lket}
\end{equation}
is the U(1) gauge invariant dipole moment basis. As shown in Appendix
A, 
\begin{equation}
\rvert\textbf{L}\rangle\equiv \begin{pmatrix} \textbf{L}_{3}, & \textbf{L}_{1}-i\textbf{L}_{2}, & \textbf{L}_{1}+i\textbf{L}_{2}, & -\textbf{L}_{3}\end{pmatrix}^{\mathrm{T}} , \label{eq:Lket2}
\end{equation}
where each $\mathbf{L}_{i}$ vector $(i=1,2,3)$ is defined in terms
of the $\hat{\textbf{d}}$ vector as 
\begin{equation}
\mathbf{L}_{i}\equiv e\sum_{\textbf{k}\in\text{BZ}}\left(i\,\partial_{\textbf{k}}\hat{d}_{i}-\epsilon_{ijl}\hat{d}_{j}\partial_{\textbf{k}}\hat{d}_{l}\right). \label{DipMomBasis}
\end{equation}
The derivation is given in Appendix~\ref{App-1}. Substituting Eq. (\ref{eq:Lket2}) and (\ref{DipMomBasis}) into Eq.
(\ref{DipMomExc}) yields the exact expression for the dipole moments
of the three bound states in terms of the components of the matrix
$\mathsf{u}_{ji}$ that diagonalizes $\hat{w}$, and the $\hat{\textbf{d}}$
vector,
\begin{equation}
\boldsymbol{\ell}_{\nu}=\frac{1}{2\sqrt{2w_{\nu}}}\sum_{j}\mathsf{u}_{j\nu}^{*}\mathbf{L}_{j}.\label{ElDipMomThree}
\end{equation}
The resulting dipole moments are hence inherently model-dependent. 

\subsubsection{Flat band non-linear sigma models }

The exciton effective dipole moment $\boldsymbol{\mathbb{\ell}}$
can be readily calculated for a family of $\hat{\textbf{d}}$ vectors
that satisfy the equation of motion of the NLsM \cite{Polyakov},
\begin{equation}
\partial_{\alpha}\hat{d}_{i}=-\tau\epsilon_{\alpha\beta}\epsilon_{ijl}\hat{d}_{j}\partial_{\beta}\hat{d}_{l},\label{eq:NLsM}
\end{equation}
where $\partial_{\alpha}\equiv\partial/\partial k_{\alpha}$, and
$\tau=\pm$ defines the winding sign (sign of the valence-band Chern
number). Eq. (\ref{eq:NLsM}) describes the minimum energy configuration
of the NLsM for a given Pontryagin index $Q=\int\frac{\text{d}^{2}\mathbf{k}}{(2\pi)^{2}}(\partial_{x}\hat{\mathbf{d}}\times\partial_{y}\hat{\mathbf{d}})\cdot\hat{\mathbf{d}}$.
$Q$ is the total skyrmion number of the unit vector $\hat{\mathbf{d}}$
when mapped from the torus to the unit sphere, and is related to the
Chern number of the bands as $\mathcal{C}=\pm Q$. Condition (\ref{eq:NLsM})
is equivalent to state that $\hat{\textbf{d}}$ can be parametrized
in terms of a single meromorphic function on the torus \cite{Polyakov}.
This family of flat band Hamiltonians was shown to saturate the lower
bound of the trace of the quantum metric \cite{Wen}. 

From Eq. (\ref{DipMomBasis}), we obtain
\begin{equation}
\mathbf{L}_{i}=\mathbf{L}_{i,0}\left(\begin{array}{c}
1\\
-i\tau
\end{array}\right),\label{eq:Li}
\end{equation}
with
\begin{equation}
\mathbf{L}_{i,0}\equiv e\sum_{\mathbf{k}\in\text{BZ}}(i\partial_{x}-\tau\partial_{y})\hat{\mathbf{d}}_{i}(\mathbf{k}).\label{eq:Li2}
\end{equation}
Therefore, the effective dipole moments of all three bound states
are circularly polarized with the same circular polarization,
\begin{equation}
\boldsymbol{\ell}_{\nu}=\boldsymbol{\ell}_{\nu,0}\left(\begin{array}{c}
1\\
-i\tau
\end{array}\right),\quad\boldsymbol{\ell}_{\nu,0}=\frac{1}{2\sqrt{2w_{\nu}}}\sum_{j}\mathsf{u}_{j\nu}^{*}\mathbf{L}_{j,0}.\label{eq:ell2}
\end{equation}
The handedness of the light polarization is fixed by the sign of the
Chern number of the bands, which is a global property of the 2D BZ. 

Next, we investigate the optical selection rules for two specific topological
models: the flattened BHZ model on a square lattice and the flattened
Haldane model on a honeycomb lattice.

\subsubsection{The flattened BHZ Model} \label{BHZ=000020model} 

The original BHZ model  \cite{Bernevig} is formed by two time reversal related blocks of $2\times2$ Hamiltonians, which
together describe the quantum spin Hall effect \cite{Kane} in topological insulators.  The single spin version of the BHZ model is a 
$2\times 2$ 
Hamiltonian corresponding to a single spin block, $\mathcal{H}_{\text{BHZ}}(\mathbf{k})=\mathbf{d}(\mathbf{k})\cdot\boldsymbol{\sigma}$,
defined on a square lattice with lattice vectors $\textbf{a}_{1}=a(1,0)$
and $\textbf{a}_{2}=a(0,1)$. The components of the $\mathbf{d}(\mathbf{k})$
vector are $d_{1}(\textbf{k})=M+t\sum_{i=x,y}\cos(k_{i}a),$ $d_{2}(\textbf{k})=-t\sin(k_{y}a)$
and $d_{3}(\textbf{k})=t\sin(k_{x}a),$ where $M$ and $t$ are real
parameters.  An important feature is that this
model has non-trivial Chern bands when the condition $0<|M/t|<2$
is satisfied. In that case, the Chern numbers of the bands are $\mathcal{C}_{\mathrm{c}}=-\mathcal{C}_{\mathrm{v}}=\mathrm{sgn}(M/t)$.
%For $M=0$ the gap closes at $\Gamma$, forming a single Dirac point. At $M=\pm2$ the gap closes 

With this $\mathbf{d}(\mathbf{k})$
vector, the dipole moment basis \eqref{DipMomBasis} for
the flattened BHZ model takes the form
\begin{equation}
\textbf{L}_{1}=0,\quad\textbf{L}_{2}=L_{0}\begin{pmatrix}1\\
0
\end{pmatrix},\quad\textbf{L}_{3}=L_{0}\begin{pmatrix}0\\
1
\end{pmatrix},
\end{equation}
where
\begin{equation}
L_{0}\equiv e\sum_{\textbf{k}\in\text{BZ}}\left(\hat{d}_{1}\partial_{k_{x}}\hat{d}_{3}-\hat{d}_{3}\partial_{k_{x}}\hat{d}_{1}\right).\label{eq:L0}
\end{equation}
The coefficients of the auxiliary matrix $\hat{w}$, defined in Eq.
(\ref{eq:ci}) and (\ref{eq:cij}) for an arbitrary unit vector $\hat{\mathbf{d}}$,
are given by $c_{1}=\delta$, $c_{2,3}=0$, and $c_{ij}=\delta_{ij}D_{i}$,
with $D_{2}=D_{3}=(1-D_{1})/2$.

By constructing the auxiliary matrix $\hat{w}$, we obtain the eigenvalues:
\begin{eqnarray}
 &  & w_{1}=\frac{1-D_{1}}{2},\quad w_{2}=\frac{1+D_{1}-2\delta}{4},\nonumber \\
 &  & w_{3}=\frac{1+D_{1}+2\delta}{2},
\end{eqnarray}
with the unitary matrix 
\begin{equation}
\hat{\mathsf{u}}=\begin{pmatrix}1 & 0 & 0\\
0 & \frac{1}{\sqrt{2}} & \frac{i}{\sqrt{2}}\\
0 & -\frac{i}{\sqrt{2}} & -\frac{1}{\sqrt{2}}
\end{pmatrix}.
\end{equation}
Substituting into Eq.~(\ref{ElDipMomThree}), we find the three dipole
moments: 
\begin{equation}
\boldsymbol{\ell}_{1}=0,\quad\boldsymbol{\ell}_{2}=\frac{\ell_{2,0}}{\sqrt{2}}\begin{pmatrix}1\\
 i
\end{pmatrix},\quad\boldsymbol{\ell}_{3}=-\frac{\ell_{3,0}}{\sqrt{2}}\begin{pmatrix}i\\
1
\end{pmatrix},\label{BHZDipMom}
\end{equation}
where $\ell_{\nu,0}=L_{0}/\sqrt{2w_{\nu}}$.
The $\nu=1$ exciton is dark, whereas the other two are bright and
couple to circularly polarized light with opposite circular polarizations.

%The approximate symmetry of the envelope function $\mathcal{R}_{\nu,\mathbf{k}}$
%$(\nu=1,2,3$) for the flattened BHZ model can be extracted in the
%vicinity of the center of the BZ in the relative momentum coordinates,
%where the Berry curvature is concentrated. 
We assume the `symmetric
gauge', in which the non-analytic part of the interaction term $W_{\mathbf{q},\mathbf{q}^{\prime}}$
is absorbed by the envelope function $\mathcal{R}_{\nu,\mathbf{k}}$,
making the latter itself invariant under $U(1)$ gauge transformations.
With this choice, $W_{\mathbf{q},\mathbf{q}^{\prime}}$ is invariant
under rotations around $\mathbf{q}=0$, whereas $\mathcal{R}_{\nu,\mathbf{k}}$
picks a topological phase factor that gives it a finite vorticity
$\zeta$ \cite{Xie}. 

\begin{figure*}
\begin{centering}
\includegraphics[scale=0.56]{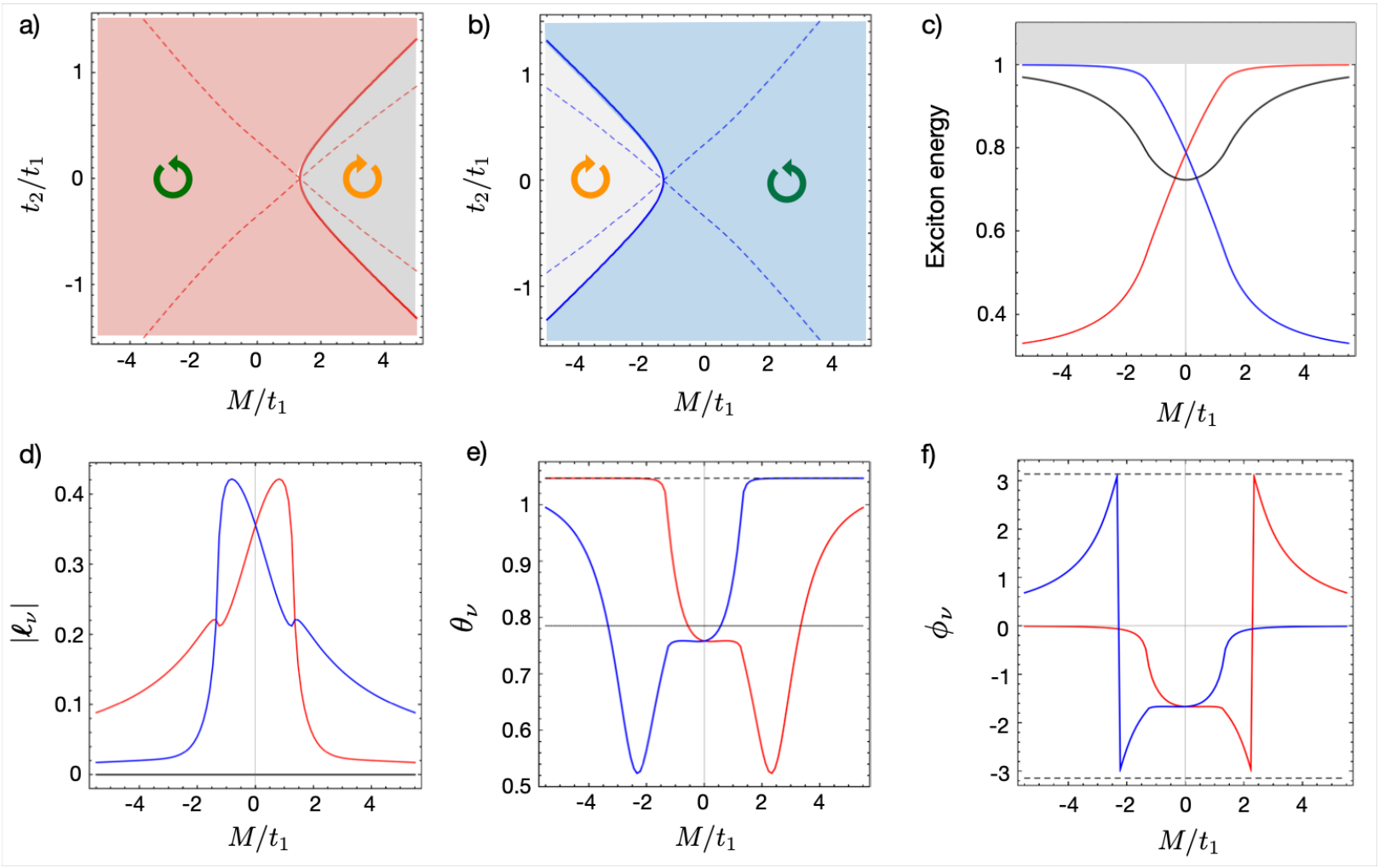}
\par\end{centering}
\caption{{\small Optical selection rules for topological excitons in the flattened
Haldane model for interaction strength $v=0.7\Delta$. Panels a) and
b) show the phase diagram describing the elliptic polarization of
light that couples to excitons $\nu=1,2$, respectively, as functions
of dimensionless parameters $t_{2}/t_{1}$ and $M/t_{1}$. The red
(blue) solid line indicates the phase boundary where $\boldsymbol{\ell}_{1(2)}$
is linearly polarized. The orange and green circles with arrows indicate
the handedness of $\boldsymbol{\ell}_{1(2)}$ in the corresponding
region. The blue (red) dashed lines indicate the phase space where
$\boldsymbol{\ell}_{1(2)}$ is circularly polarized. We take $t_{2}/t_{1}=0.5$
to obtain panels c)$-$f), where the red, blue and black curves describe
excitons $\nu=1,2,3$, respectively, where $\nu=1,2$ are bright and
$\nu=3$ is dark. c) Energy of the three exciton bound states as functions
of $M/t_{1}$. The shaded region indicates the single-particle states
continuum, at the band edge. d) Effective exciton dipole moment $|\boldsymbol{\ell}_{\nu}|$,
e) polarization angle $\theta_{\nu}$ and polarization angle $\phi_{\nu}$
in radians as functions of $M/t_{1}$. The dashed and solid horizontal
lines in e) indicate $\theta=\pi/3$ and $\theta=\pi/4$, respectively.
The dashed lines in panel f) indicate $\phi=\pm\pi$. }} 
\end{figure*}

We show in Fig.~\ref{fig2}(a) the envelope function of exciton $\nu=2$ in the BZ  for the interaction
strength $v=0.7\Delta$ and $M/t=1$. 
The vectors shown in the plot represent the complex phase of $\mathcal{R}_{\mathbf{k},2}$, with the amplitude being indicated in the color bar. 
The three excitons have the same vorticity $\zeta=2$. The profile functions of excitons $\nu=1$ and 3 are shown in Appendix B. In Fig.~\ref{fig2}(b,c),  the black  lines correspond to
a dark exciton $\nu=1$, whereas the red ($\nu=2$) and blue ($\nu=3$)
lines correspond to bright excitons. The handedness of the effective dipole moments $\boldsymbol{\ell}_{2,3}$ is indicated in the blue and red circles, respectively. The evolution of the energy spectrum of the excitons
and the magnitude of the effective exciton dipole moment $|\boldsymbol{\ell}_\nu|$ as a function of the
ratio $M/t$ are shown in Fig.~\ref{fig2}(b) and (c), respectively.

\subsubsection{The flattened Haldane Model}

The $\textbf{d}$-vector in the Haldane model is given by $d_{1}(\textbf{k})=t_{1}\sum_{j}\cos(\textbf{k}\cdot\textbf{a}_{j}),$
$d_{2}(\textbf{k})=t_{1}\sum_{j}\sin(\textbf{k}\cdot\textbf{a}_{j})$,
and $d_{3}(\textbf{k})=M-t_{2}\sum_{j}\sin(\textbf{k}\cdot\textbf{b}_{j})$,
where $t_{1},t_{2}$ are real hopping amplitudes for first and second
nearest neighbors, respectively, in the honeycomb lattice and $M$
is a topologically trivial gap \cite{Haldane}. This lattice has three
nearest neighbor vectors $\textbf{a}_{1}=a(1,0)$, $\textbf{a}_{2}=a(-1/2,\sqrt{3}/2)$
and $\textbf{a}_{3}=-\textbf{a}_{1}-\textbf{a}_{2}$, and six second
nearest neighbor vectors, $\mathbf{b}_{1}=\mathbf{a}_{2}-\mathbf{a}_{3}$,
$\mathbf{b}_{2}=\mathbf{a}_{3}-\mathbf{a}_{1}$, and so on. For simplicity,
we assume particle-hole symmetry, in which case the model is topological
when $M/t_{2}<3\sqrt{3}$, with Chern numbers $\mathcal{C}=\pm1$. 

The coefficients of the auxiliary matrix $\hat{w}$ in Eq. (\ref{AuxMat-1})$-$(\ref{eq:cij})
are specified as $c_{1,2}=0$, $c_{3}=\delta$, and $c_{\alpha\beta}=\delta_{\alpha\beta}D_{\alpha}$,
with $D_{1}=D_{2}$ and $D_{3}=1-2D_{1}$. The resulting eigenvalues
of $\hat{w}$ are 
\begin{eqnarray}
 &  & w_{1}=\frac{1-D_{1}-\delta}{2},\quad w_{2}=\frac{1-D_{1}+\delta}{2},\nonumber \\
 &  & w_{3}=D_{1},\label{eq:w2}
\end{eqnarray}
with the corresponding unitary matrix 
\begin{equation}
\hat{\mathsf{u}}=\begin{pmatrix}\frac{1}{\sqrt{2}} & \frac{i}{\sqrt{2}} & 0\\
-\frac{i}{\sqrt{2}} & -\frac{1}{\sqrt{2}} & 0\\
0 & 0 & 1
\end{pmatrix}.\label{eq:u4}
\end{equation}
The envelope function solutions $\mathcal{R}_{\nu,\mathbf{k}}$
($\nu=1,2,3$) of this model have been previously calculated in Ref.
\cite{Xie}, and have vorticity $\zeta=2$. 

From Eq. (\ref{ElDipMomThree}), (\ref{eq:w2}) and (\ref{eq:u4}),
the three dipole moments can be cast in the form
\begin{equation}
\boldsymbol{\ell}_{1}=\frac{\textbf{L}_{1}+i\textbf{L}_{2}}{4\sqrt{w_{1}}},\quad\boldsymbol{\ell}_{2}=-\frac{i\textbf{L}_{1}+\textbf{L}_{2}}{4\sqrt{w_{2}}},\quad\boldsymbol{\ell}_{3}=\frac{\textbf{L}_{3}}{2\sqrt{2w_{3}}},\label{eq:LHM}
\end{equation}
where $\mathbf{L}_{\nu=1,2,3}$ are defined in Eq.~(\ref{DipMomBasis}). In a more explicit form, we find that $\boldsymbol{\ell}_{3}=0$
(dark exciton), whereas $\boldsymbol{\ell}_{1,2}$ can be represented
by the Jones vector for elliptic polarization, 
\begin{equation}
\boldsymbol{\ell}_{\nu}=|\boldsymbol{\ell}_{\nu}|\left(\begin{array}{c}
\cos\theta_{\nu}\\
\sin\theta_{\nu}\text{e}^{i\phi_{\nu}}
\end{array}\right).\label{eq:l4}
\end{equation}
Analytical expressions for Eq.~(\ref{eq:l4}) are derived in Appendix~\ref{App-1}. 

We show the phase diagram of the polarization of $\mathbf{\boldsymbol{\ell}}_{1(2)}$
versus $M/t_{1}$ and $t_{2}/t_{1}$ in panels a (b) of Fig. 3 for
interaction strength $v=0.7\Delta$. The handedness of $\mathbf{\boldsymbol{\ell}}_{\nu}$
is $\text{sgn}(\phi_{\nu})=\pm1$ for clockwise and counterclockwise
polarizations, which are indicated by the circles. The solid red (blue)
line indicates the boundary between regions with opposite handedness,
where $\mathbf{\boldsymbol{\ell}}_{1(2)}$ is linearly polarized.
Dashed (blue) curves indicate points where $\mathbf{\boldsymbol{\ell}}_{1(2)}$
is circularly polarized. Panels 3c$-$f correspond to $t_{2}/t_{1}=0.5$.
Fig. 3c shows the exciton energy $\varepsilon_{\nu}$ as functions
of $M/t_{1}$. The other panels, d)$-$f), show the amplitude of the
effective exciton dipole moment $|\boldsymbol{\ell}_{\nu}|$, and
the polarization angles $\theta_{\nu}$ and $\phi_{\nu}$ versus $M/t_{1}$. 

\begin{figure*}
\begin{centering}
\includegraphics[scale=0.5]{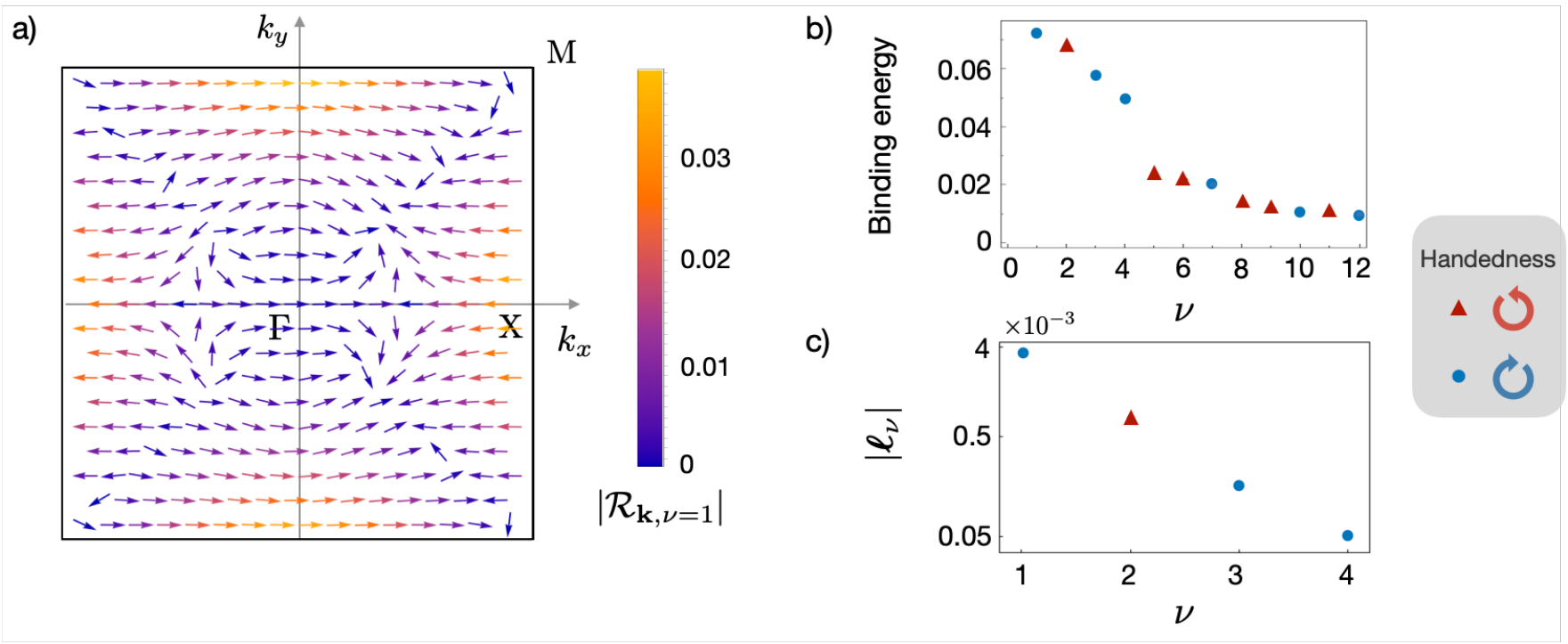}
\par\end{centering}
\caption{{\small Topological excitons for the flattened BHZ model with Coulomb
interactions. a) Profile function $\mathcal{R}_{\mathbf{k},\nu}(0)$
of the lowest energy state $\nu=1$. Arrows indicate the complex phase
of $\mathcal{R}_{\mathbf{k}}$ and the magnitude is represented in
the color bar. All exciton bound states have the same vorticity $\zeta=2$
and couple to circularly polarized light. b) Normalized binding energy
$1-\varepsilon_{\nu}/\Delta$  numerically calculated
for the first twelve exciton states $\nu=1,2,\ldots,12$. c) Effective
exciton dipole moment amplitude $|\boldsymbol{\ell}_{\nu}|$ (in units of $ea$) for the four lowest energy excitons. Vertical axis is shown in logarithmic scale. The
linear response exciton oscillation strength $o_{\nu,\sigma=+1}=|\boldsymbol{\ell}_{\nu}|^{2}$
decreases exponentially with increasing $\nu$. In panels b) and c)
the handedness of the effective exciton dipole moment is indicated
by red triangles and blue dots.}}
\end{figure*}

\subsection{Coulomb interaction}

Coulomb interactions generate an infinite number of excitonic bound
states inside the insulating gap, $\nu=1,2\ldots,\infty$. The interaction $v(\textbf{k})$
takes the form of the RK potential \cite{Rytova,Keldysh}
\begin{equation}
v(\mathbf{q})=\frac{A_{0}}{(2\pi)^{2}}\int V_{\text{HK}}(\mathbf{r})\text{e}^{-i\textbf{q}\cdot\textbf{r}}\,d\textbf{r},\label{eq:v(q)}
\end{equation}
where
\begin{equation}
V_{\text{RK}}(\mathbf{r})=\frac{e^{2}}{\epsilon r_{0}}\left[H_{0}(r/r_{0})-Y_{0}(r/r_{0})\right].\label{eq:Vrk}
\end{equation}
$r_{0}$ is an effective screening length, $\epsilon$ is the high-frequency
dielectric constant, $e$ is the electron charge, and $H_{0}(x)$
and $Y_{0}(x)$ are the Struve and Neumann functions, respectively.
The RK potential has the following asymptotic behavior, $V_{\mathrm{RK}}(\textbf{r})\sim1/r$
for $r\gg r_{0}$ and $V(\textbf{r})\sim-\ln(r/r_{0})$ for $r\ll r_{0}$.

We performed a numerical calculation of the solution of the Wannier
Eq. (\ref{eq:Wannier-1}) for the flattened BHZ model by discretizing
the BZ into a $70\times70$ grid. We then numerically obtained the
exciton eigenenergies and eigenvectors by diagonalizing this matrix.
In Fig. 4a we show the envelope function of the lowest energy state
$\mathcal{R}_{\mathbf{k},1}$, which has the same vorticity $\zeta=2$  as in the short range case. 
Higher energy excitons have distinct envelope functions (see Appendix B), although the vorticity  remains
the same for all states. In Fig. 4b, we show the normalized
binding energy $1-\varepsilon_{\nu}/\Delta$ of the twelve lowest energy
states. All excitons calculated
are bright and couple with circularly polarized light. The handedness
of the polarization of $\boldsymbol{\ell}_{\nu}$ is indicated by
the blue circles and red triangles. The ground state ($\nu=1$) and the $\nu=3,4,7,10,12$ states 
have the same handedness, whereas the $\nu=2,5,6,8,9,11$ states have the opposite 
circular polarization. In Fig. 4c we show the the amplitude of the effective dipole moment
for the first four exciton states $\nu=1,2,3,4$. 
$|\boldsymbol{\ell}_{\nu}|$ scales exponentially with the exciton
index $\nu$, with the $\nu=1$ state being the brightest. Hence,
higher energy excitons are significantly dimmer in comparison with
the $\nu=1$ state. 

\section{Conclusion} \label{sec-4}

Selection rules have been successfully used to describe optical transitions
in excitonic states of conventional insulators \cite{Elliot}. They have also been successfully modified to 
account for the angular momentum contribution due to winding numbers that are associated to the chirality of 2D massive Dirac fermions  \cite{Xiao1,Wang-1,Zhang,Cao-1,Gong,Ju,Park}. Selection
rules for excitonic optical transitions in flat bands on the other
hand are governed by quantum geometric effects in combination with
possible topological constraints. In this paper, we addressed how
global properties of the 2D BZ will affect the polarizability of excitons
in flat bands.

We derived the optical selection rules for topological excitons considering
both short range and Coulomb interactions. The former case can be
addressed exactly by constructing an auxiliary matrix formalism. We then obtained analytic expressions
for the effective exciton dipole moment in a family of flat two-band
Hamiltonians with skyrmion pseudo-spin texture (NLsM),  in
the flattened BHZ model and also in the flattened Haldane model. We showed that in
the NLsM three bright topological excitons emerge in the insulating
gap, each one coupling with the same circular polarization of light
that is fixed by the Chern number of the electronic bands. For the
flattened versions of the BHZ and Haldane models with short range
interactions, only two excitons are bright, with the third being dark.
The bright excitons in the flattened BHZ model couple to circularly
polarized light with opposite handedness, whereas in the later the
excitons couple to elliptically polarized light with also opposite handedness. 

Finally, we numerically calculated the non-hydrogenic spectrum of
topological excitons in the flattened BHZ model for Coulomb interactions. We showed that the effective
exciton dipole moment is finite for all calculated excitons, with no dark excitons being present, and circularly
polarized. We calculated the oscillation strength of the excitons,
which indicates their brightness, and showed that it decays exponentially
with the exciton index $\nu$. 

We have established though explicit calculations in different models
that topological excitons created with linearly polarized light in
flat bands have an effective dipole moment that selectively couples
with a given circular or elliptic polarization of light. Even though
the envelope function has a fixed vorticity that is determined
by the topology of the electronic bands, the polarization of the effective
exciton dipole moment is model dependent and is indicated by the optical
selection rule. Our results highlight how band topology influences
exciton properties, providing a way for predicting light-matter coupling
in topological materials with flat bands. 
\begin{acknowledgments}
BU acknowledges NSF grant DMR-2529526 for support. The work of H.-Y.X. is supported by the Dodge
Family Fellowship granted by the University of Oklahoma. 
\end{acknowledgments}

\appendix

\section{Effective exciton dipole moment} \label{App-1}

The dipole momentum ket basis defined in Eq. (\ref{eq:Lket}) is $\rvert\textbf{L}\rangle\equiv4e\sum_{\mathbf{k}\in\text{BZ}}\boldsymbol{\mathcal{A}}_{\text{cv}}(\mathbf{k})|\mathbf{k},0\rangle$.
The unitary transformation that diagonalizes the generic flat band
Hamiltonian $\hat{\mathcal{K}}_{\mathbf{k}}=\frac{\Delta}{2}\hat{\mathbf{d}}(\mathbf{k})\cdot\boldsymbol{\sigma}$
is 
\begin{equation}
\hat{U}(\mathbf{k})=\left(\begin{array}{cc}
\sqrt{\frac{1+\hat{d}_{3}(\mathbf{k})}{2}} & -\sqrt{\frac{1-\hat{d}_{3}(\mathbf{k})}{2}}\\
\frac{\hat{d}_{\parallel}(\mathbf{k})}{\sqrt{2+2\hat{d}_{3}(\mathbf{k})}} & \frac{\hat{d}_{\parallel}(\mathbf{k})}{\sqrt{2-2\hat{d}_{3}(\mathbf{k})}}
\end{array}\right),\label{eq:U2}
\end{equation}
where $\hat{\mathbf{d}}(\mathbf{k})=(\hat{d}_{1},\hat{d}_{2},\hat{d}_{3})$
is a unit vector, with $\sum_{i}\hat{d}_{i}^{2}=1$, and $\hat{d}_{\parallel}\equiv\hat{d}_{1}+i\hat{d}_{2}$.
From Eq. (\ref{eq:ehproduct}), the particle hole-pair state is 
\begin{equation}
|\mathbf{k},0\rangle=\left(-\frac{|\hat{d}_{\parallel}|}{2},\frac{(1+\hat{d}_{3})\hat{d}_{\parallel}^{*}}{2|\hat{d}_{\parallel}|},-\frac{(1-\hat{d}_{3})\hat{d}_{\parallel}}{2|\hat{d}_{\parallel}|},\frac{|\hat{d}_{\parallel}|}{2}\right)^{T}.\label{eq:K0}
\end{equation}
The interband Berry connection, defined in Eq. (\ref{eq:P-2}), that
corresponds to the unitary transformation (\ref{eq:U2}) is
\begin{align}
\boldsymbol{\mathcal{A}}_{\text{cv}}(\mathbf{k}) & =-\frac{i}{2}\frac{\hat{d}_{\parallel}^{*}\partial\hat{d}_{\parallel}+(1+\hat{d}_{3})\partial_{\mathbf{k}}\hat{d}_{3}}{|\hat{d}_{\parallel}|}\nonumber \\
 & =\frac{i}{2}\frac{\hat{d}_{\parallel}\partial_{\mathbf{k}}\hat{d}_{\parallel}^{*}-(1-\hat{d}_{3})\partial_{\mathbf{k}}\hat{d}_{3}}{|\hat{d}_{\parallel}|}.\label{eq:A3}
\end{align}
Combining Eq. (\ref{eq:K0}) and (\ref{eq:A3}) with Eq. (\ref{eq:Lket}),
we arrive at Eq. (\ref{eq:Lket2}) and (\ref{DipMomBasis}). 

In the basis of the reciprocal vectors $\mathbf{K}_{1,2}$, a wave
vector is decomposed into $\mathbf{k}=x_{1}\frac{\mathbf{K}_{1}}{2\pi}+x_{2}\frac{\mathbf{K}_{2}}{2\pi}$
with $x_{1,2}\in[0,2\pi)$, which can be written in the matrix form
$\mathbf{k}=\hat{O}^{T}\boldsymbol{x}$, where $\boldsymbol{x}=(x_{1},\,x_{2})^{T}$
and 
\begin{equation}
\hat{O}=\frac{1}{2\pi}\left(\begin{array}{cc}
K_{1,x} & K_{1,y}\\
K_{2,y} & K_{2,y}
\end{array}\right).\label{eq:O}
\end{equation}
The infinitesimal BZ area and the gradient operator transform as $\text{d}k_{x}\text{d}k_{y}=|\mathrm{Det}\hat{O}|\text{d}x_{1}\text{d}x_{2}=\frac{A_{\mathrm{BZ}}}{(2\pi)^{2}}\text{d}x_{1}\text{d}x_{2}$
and $\partial_{\mathbf{k}}=\hat{O}^{-1}\partial_{\boldsymbol{x}}$,
respectively, where $A_{\text{BZ}}$ is the area of the BZ. Therefore,
the dipole moment basis in Eq.~(\ref{DipMomBasis}) take the expression
$\mathbf{L}_{i}=e\hat{O}^{-1}\tilde{\mathbf{L}}$, where
\begin{align}
\tilde{\mathbf{L}}_{i}= & \frac{i}{(2\pi)^{2}}\int_{0}^{2\pi}\text{d}x\left(\begin{array}{c}
\hat{\tilde{d}}_{i}(2\pi,x)-\hat{\tilde{d}}_{i}(0,x)\\
\hat{\tilde{d}}_{i}(x,2\pi)-\hat{\tilde{d}}_{i}(x,0)
\end{array}\right) \nonumber \\
 & -\int\negmedspace\int_{0}^{2\pi}\frac{\text{d}\boldsymbol{x}}{(2\pi)^{2}}\epsilon_{ijl}\hat{\tilde{d}}_{j}\partial_{\boldsymbol{x}}\hat{\tilde{d}}_{l},
\end{align}
with $\tilde{\mathbf{d}}(\boldsymbol{x})\equiv\mathbf{d}(\mathbf{k}(\boldsymbol{x}))$.
This expression gives the dipole moment basis in Eq. (\ref{BHZDipMom})
for the flattened BHZ model.

\begin{figure*}
\begin{centering}
\includegraphics[scale=0.47]{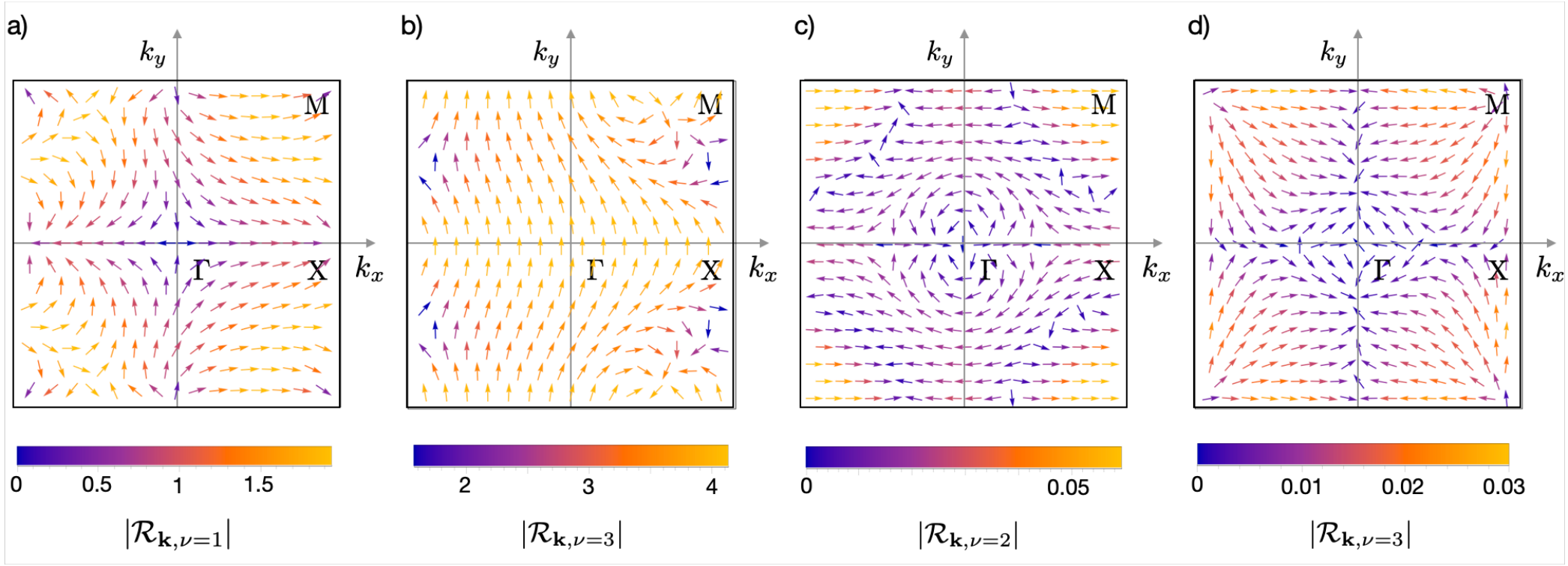}
\par\end{centering}
\caption{{\small Profile functions $\mathcal{R}_{\mathbf{k},\nu}$ for the BHZ model. a) Dark exciton $\nu=1$ and b) bright exciton $\nu=3$ for short range interactions $v=0.7\Delta$ and $M/t=1$. Panels c) and d) show the profile functions for the case of Coulomb interactions. c) first excited state ($\nu=2$). d) second excited state ($\nu=3$). }} \label{fig5}
\end{figure*}

For the flattened Haldane model we obtain
\begin{equation}
\tilde{\mathbf{L}}_{1}=\left(\begin{array}{c}
R_{1}\\
-R_{1}
\end{array}\right),\quad\tilde{\mathbf{L}}_{2}=\left(\begin{array}{c}
R_{2}+iI_{2}\\
R_{2}-iI_{2}
\end{array}\right),\quad\tilde{\mathbf{L}}_{3}=0,\label{eq:Ltilde}
\end{equation}
where
\begin{align}
R_{i=1,2}(M,t_{2}) & =-\int_{0}^{2\pi}\frac{\text{d}^{2}\mathbf{x}}{(2\pi)^{2}}\epsilon_{ijl}\hat{\tilde{d}}_{j}\partial_{\boldsymbol{x}}\hat{\tilde{d}}_{l}\nonumber \\
I_{2}(M,t_{2}) & =\int_{0}^{2\pi}\frac{\text{d}x}{(2\pi)^{2}}\left[\hat{\tilde{d}}_{2}(2\pi,x)-\hat{\tilde{d}}_{2}(0,x)\right].\label{eq:R}
\end{align}
We assume that $M$ and $t_{2}$ are normalized by $t_{1}$. It is
readily to verify the symmetry properties $R_{1}(M,-t_{2})=-R_{1}(-M,t_{2})=R_{1}(M,t_{2})$,
$R_{2}(-M,t_{2})=-R_{2}(M,-t_{2})=R_{2}(M,t_{2})$, and $I_{2}(-M)=I_{2}(M)$.
In Eq. (61), we can prove that $D_{1}(-M,t)=D_{1}(M,-t)=D_{1}(M,t)$
and $\delta(M,-t_{2})=-\delta(-M,t_{2})=\delta(M,t_{2})$. The dipole
moment basis reads
\begin{equation}
\mathbf{L}_{1}=ea\left(\begin{array}{c}
\frac{3}{2}R_{1}\\
\frac{3\sqrt{3}}{2}R_{1}
\end{array}\right),\quad\mathbf{L}_{2}=ea\left(\begin{array}{c}
-\frac{3}{2}R_{2}+i\frac{3}{2}I_{2}\\
\frac{3\sqrt{3}}{2}R_{2}+i\frac{3\sqrt{3}}{2}I_{2}
\end{array}\right),\label{eq:L5}
\end{equation}
and $\mathbf{L}_{3}=0.$ From Eq.~(\ref{eq:l4}) we obtain the effective
exciton dipole moments
\begin{align}
\boldsymbol{\ell}_{1} & =\frac{ea}{8\sqrt{w_{1}}}\left(\begin{array}{c}
3(R_{1}-I_{2})-3iR_{2}\\
3\sqrt{3}(R_{1}-I_{2})+\sqrt{3}iR_{2}
\end{array}\right),\nonumber \\
\boldsymbol{\ell}_{2} & =\frac{ea}{8\sqrt{w_{2}}}\left(\begin{array}{c}
3R_{2}-3i(R_{1}+I_{2})\\
-\sqrt{3}R_{2}-3\sqrt{3}i(R_{1}+I_{2})
\end{array}\right),\label{eq:l6}\\
\boldsymbol{\ell}_{3} & =0.\nonumber 
\end{align}
In the Jones vector representation for $\boldsymbol{\ell}_{1,2}$,
\[
\boldsymbol{\ell}_{\nu}=|\boldsymbol{\ell}_{\nu}|\left(\begin{array}{c}
\cos\theta_{\nu}\\
\sin\theta_{\nu}\text{e}^{i\phi_{\nu}}
\end{array}\right),
\]
we have
\begin{align}
|\boldsymbol{\ell}_{\nu}| & =ea\frac{\sqrt{3R_{2}^{2}+9(R_{1}+\sigma_{\nu}I_{2})^{2}}}{4\sqrt{w_{\nu}}}\nonumber \\
\theta_{\nu} & =\tan^{-1}\sqrt{\frac{R_{2}^{2}+9(R_{1}+\sigma_{\nu}I_{2})^{2}}{3R_{2}^{2}+3(R_{1}+\sigma_{\nu}I_{2})^{2}}}\label{eq:=00005Ctheta2}\\
\phi_{\nu} & =\mathrm{Arg}\big( 3 (I_2-\sigma_\nu R_1)^2 - R_2^2 - i 4 R_2  (I_2-\sigma_\nu R_1) \big),\nonumber 
\end{align}
where $\sigma_{1}=-1$ and $\sigma_{2}=+1$ and $\mathrm{Arg}(z)$ is the argument of a complex number $z$. 

In general, $\boldsymbol{\ell}_{1,2}$ are elliptically polarized
and the polarization handedness of $\boldsymbol{\ell}_{1}$ is $-\mathrm{sgn}[R_{2}(I_{1}-R_{2})]$
and that of $\boldsymbol{\ell}_{2}$ is $-\mathrm{sgn}[R_{2}(I_{1}+R_{2})]$.
We note three special cases. \emph{i)} When $\left|R_{2}/(R_{1}\mp I_{2})\right|=\sqrt{3}$,
$\boldsymbol{\ell}_{1,2}$ is circularly polarized.\emph{ ii)} When
$R_{2}=0$, $\boldsymbol{\ell}_{1,2}\propto\left(1,\sqrt{3}\right),$
which is linearly polarized. \emph{iii)} When $R_{1}=\pm I_{2}$,
$\boldsymbol{\ell}_{1(2)}\propto\left(1,-\sqrt{3}\right)$, which
is also linearly polarized. Those three cases are highlighted through
the solid and dashed lines in the phase diagrams of panels a) and
b) in Fig. 3.

\section{Profile functions in the BHZ model}

The profile function of the bright $\nu=2$ exciton in the BHZ model with short range interactions is shown in Fig. 2 of the main text.  In Fig. 5a, b we show the profile functions of the other two exciton states, $\nu=1$ (dark) and $\nu=3$ (bright). As anticipated, all excitons share the same vorticity $\zeta=2$, which is dictated by the topology of the electronic bands.  In panels 5c and d we show the profile functions of the first and second excited excitons in the non-hydrogenic spectrum of the BHZ model in the presence of Coulomb interactions.

\end{document}